# Modelling small block aperture in an in-house developed GPU-accelerated Monte Carlo-based dose engine for pencil beam scanning proton therapy

Running title: block aperture in VPMC

Hongying Feng[1], PhD, Jason M. Holmes[1], PhD, Sujay A. Vora, MD, Joshua B. Stoker, PhD, Martin Bues, PhD[1], William W. Wong, MD, Terence S. Sio, MD, MS[1], Robert L. Foote, MD[2], Samir H. Patel, MD[1], Jiajian Shen[1], PhD, Wei Liu[1], PhD

[1]Department of Radiation Oncology, Mayo Clinic, Phoenix, AZ 85054, USA

[2]Department of Radiation Oncology, Mayo Clinic, Rochester, MN 55902, USA

Corresponding author: Wei Liu, Liu.Wei@mayo.edu.

Author responsible for statistical analysis: Wei Liu, Liu.Wei@mayo.edu.

**Conflict of Interest Statement**

Conflict of Interest: None

**Funding Statement**

This research was supported by the National Cancer Institute (NCI) Career Developmental Award K25CA168984, Arizona Biomedical Research Commission Investigator Award, the Lawrence W. and Marilyn W. Matteson Fund for Cancer Research, and the Kemper Marley Foundation.

**Data Availability Statement**

Research data are stored in an institutional repository and will be shared upon request to the corresponding author.

**Abstract**

**Purpose:** To enhance an in-house graphic-processing-unit (GPU) accelerated virtual particle (VP)-based Monte Carlo (MC) proton dose engine (VPMC) to model aperture blocks in both dose calculation and optimization for pencil beam scanning proton therapy (PBSPT)-based stereotactic radiosurgery (SRS).

**Methods and Materials:** A module to simulate VPs passing through patient-specific aperture blocks was developed and integrated in VPMC based on simulation results of realistic particles (primary protons and their secondaries). To validate the aperture block module, VPMC was first validated by an opensource MC code, MCsquare, in eight water phantom simulations with 3cm thick brass apertures: four were with aperture openings of 1, 2, 3, and 4cm without a range shifter, while the other four were with same aperture opening configurations with a range shifter of 45mm water equivalent thickness. Then, VPMC was benchmarked with MCsquare and RayStation MC for 10 patients with small targets (average volume 8.4 cc with range of 0.4 - 43.3 cc). Finally, 3 typical patients were selected for robust optimization with aperture blocks using VPMC.

**Results:** In the water phantoms, 3D gamma passing rate (2%/2mm/10%) between VPMC and MCsquare was 99.71±0.23%. In the patient geometries, 3D gamma passing rates (3%/2mm/10%) between VPMC/MCsquare and RayStation MC were 97.79±2.21%/97.78±1.97%, respectively. Meanwhile, the calculation time was drastically decreased from 112.45±114.08 seconds (MCsquare) to 8.20±6.42 seconds (VPMC) with the same statistical uncertainties of ~0.5%. The robustly optimized plans met all the dose-volume-constraints (DVCs) for the targets and OARs per our institutional protocols. The mean calculation time for 13 influence matrices in robust optimization by VPMC was 41.6 seconds and the subsequent on-the-fly "trial-and-error" optimization procedure took only 71.4 seconds on average for the selected three patients.

**Conclusion:** VPMC has been successfully enhanced to model aperture blocks in dose calculation and optimization for the PBSPT-based SRS.

**Introduction**

Stereotactic radiosurgery (SRS) is commonly used for primary and metastatic brain tumors[1,2]. The conformal and compact dose distributions from SRS allows for sparing of the adjacent normal tissues with anatomical complexity in brain. Conventional SRS was first proposed and continuously developed with photon therapy. Compared to photon therapy, proton therapy has the intrinsic geometrical sparing characteristic such that a proton beam deposits all of its energy within a certain range and none beyond, known as the Bragg-peak[3-12]. Previously, proton-based SRS was proposed, using the passive scattering (PS) delivery technique[13,14]. With the development of the pencil beam scanning (PBS) delivery technique, where for each beamlet the stopping location can be steered by the magnets and the intensity can be modulated by inverse optimization[15-24], the PBS proton therapy (PBSPT) can achieve an optimal dose distribution with superior target dose conformity and adjacent organs-at-risk (OAR) sparing. A few exploratory investigations[25-27] on the PBSPT-based SRS have been conducted to pursue superior dosimetric outcomes.

Due to the beamlet level flexibility, the use of patient-specific apertures is largely eliminated in PBSPT compared to PS proton therapy, leaving the lateral fall-off of a dose distribution dominantly determined by the characteristics of individual beamlets, of which the in-air spot size ($\sigma$) decreases with increased energy. Modern PBSPT systems worldwide mostly have small spot sizes. The system at Mayo Clinic Arizona, for instance, has in-air spot sizes ($\sigma$) of approximately 5 mm to 2 mm with proton energy from 71.3 MeV to 228.8 MeV, respectively. However, brain tumors at shallow locations necessitates the use of range shifters to pull back the Bragg-peaks of proton beams. Consequently, beamlets undergo additional Coulomb scattering due to the range shifter, resulting in enlarged spot sizes possibly exceeding 15 mm[28-31]. With larger spot sizes, the lateral dose fall-off impairs the sparing of adjacent OARs[32,33]. To address this issue, a patient-specific aperture was proposed in PBSPT[34-38]. It was found that the use of patient-specific aperture is advantageous in improving lateral dose conformity and reducing the out-of-field dose and possible subsequent toxicities.

When an aperture is present in the field, the edge-scattered protons will introduce additional perturbations and possible contaminations[38,39] in the resulting dose distribution. Such contaminations are not well modelled by analytical dose calculation algorithms[40-42], which may lead to unexpected clinical results[43]. Therefore, Monte Carlo (MC)-based proton dose engines are more preferred to accurately model the apertures. RayStation MC (RaySearch Laboratories AB, Stockholm, Sweden) is a commercial treatment planning system (TPS), which supports MC-based modelling of aperture blocks[38,44]. However, the "voxelization" method used for modelling the aperture opening within RayStation MC may become problematic when the aperture opening is small, approaching the size of voxels in the proton dose calculation[42]. In a recent study[42], an independent open-source fast MC proton dose engine, MCsquare[45], was enhanced to model patient-specific aperture blocks with openings as small as 1cm across, where the aperture openings were better modelled by directly connecting the aperture opening boundary points. The precision of the enhanced MCsquare was validated to be adequate for the use of dose calculations, however, the calculation speed of the enhanced MCsquare was not fast enough for robust optimizations[20,46-55] in PBSPT treatment planning.

To achieve fast and accurate MC-based robust optimization[18,21,22,56-64] for PBSPT, we have developed a graphic processing unit (GPU)-accelerated MC dose calculation engine based on the proposed novel concept of virtual particles (henceforth referred as Virtual Particle Monte Carlo or VPMC)[65]. In this study, our aim was to enhance VPMC to model aperture blocks in both dose calculation and robust optimization for the PBSPT-based SRS treatment planning.

**Materials and Methods**

**A. Virtual Particle Monte Carlo (VPMC)**

Virtual Particle (VP) is a novel concept where the physics of protons and subsequent secondaries produced during MC-based proton therapy dose calculations, is fully reproduced by singular GPU-friendly, proton-like virtual particles. VP is a statistical concept that works by converting the histories of realistic particles (i.e., primary and secondary protons with further simplifications) to the histories of VPs, in terms of particle transport and energy deposit.

In a conventional MC simulation, primary protons are initialized at the beginning while secondary protons are generated at subsequent steps according to probability distribution functions (PDFs). However, in a VPMC simulation, all VPs are initialized at the beginning, where each VP corresponds to one proton (either primary or secondary), and the complexity and randomness of secondary proton generation is thus eliminated. As a result, the governing models and controlling logic are simplified to be identical for every VP in a VPMC simulation. Such simplifications are ideal for CUDA-parallelization, since the thread divergence problem due to the generation of secondary particles and bifurcation within a warp can be successfully circumvented.

Pre-calculated physics parameters (the deposited energy, energy straggling, the deflection angle, and the ionization probability) databases (i.e., PDFs), generated based on the simulation records of realistic particles, are used to increase the calculation efficiency in VPMC simulations, by using the efficient database querying based on CUDA texture instead of on-the-fly calculation.

Please refer to Shan et al.[65] for more details of the VP concept and the implementation of the VPMC proton dose engine.

**B. Enhancement of VPMC to Model Aperture blocks**

The aperture block is a brass block of 30 mm thickness with a patient-specific opening. The brass block with such thickness is capable of blocking protons with energy as large as about 155 MeV, which exceeds the energies clinically used for the treatment of shallow brain cancers. Dedicated applicators for proton SRS were designed, which attaches to the proton nozzle and can hold brass apertures to collimate proton beam by extending the downstream face of the aperture at 150 mm to isocenter[66]. To treat shallow tumors, a circular ABS Resin with 6.1 cm diameter and water equivalent thickness (WET) 45 mm can be placed upstream against the brass aperture to work as a range shifter[28] with its downstream face at 190 mm to isocenter (therefore a 10 mm gap between the range shifter and the aperture block, Figure 1(a)).

In the VPMC pre-calculated PDFs were only dependent on VP energy when VPs were tracked in the range shifter due to the homogenous material of the range shifter. In addition, a set of beam-line parameters was fine-tuned for accurate dose calculation based on the pre-calculated PDFs. Similar to the range shifter implementation, the aperture block was incorporated into VPMC using the same methodology except for a few aperture-specific arrangements as follows.

To account for the aperture block material in VPMC, the open-source fast MC code MCsquare was used to generate the PDFs needed for VPMC by irradiating a brass phantom with proton beams of different energies (0-230 MeV) with 20M primary protons. The histories were recorded and used to obtain PDFs describing the relationship between physics parameters $X$ and the energy E, i.e., $X(E)$, specific to the process where particles were travelling in the brass block.

Accordingly, a set of beam-line parameters was fine-tuned as well based on the derived physics parameter PDFs. For the aperture block opening, when a VP was travelling between the entrance and exit of the aperture block along the beam axis dimension, the so-called crossing number algorithm[67], also known as the ray casting algorithm, was used to determine whether the VP was in the opening or brass block for each MC simulation step. If the VP was in the opening, it would simply be transported with the previous velocity and direction. If the VP was in the brass block, the status of the VP would be updated according to the physics parameters sampled from the pre-generated aperture block-related physics parameter PDFs.

Please note that MCsquare has been thoroughly validated against other MC codes and measurements in both phantoms and patient geometries[41,42,45,68-71]. In addition, MCsquare has been fully commissioned and has been incorporated into our in-house TPS[23,24,68,72], and has been clinically used as the second monitor unit (MU) check system at our proton center for years[42,68] and other proton centers[73]. More importantly, MCsquare has been enhanced to model the aperture block and validated with measurements and RayStation MC[42]. Therefore, we may safely assume that the aperture enhanced MCsquare is a good anchor for the development and validation of aperture enhanced VPMC in terms of accuracy.

**C. Implementation of highly efficient optimization with low memory usage**

The newly enhanced VPMC, capable of modelling apertures, has been integrated in our in-house TPS, named Shiva[23,24,72,74-76], which was implemented using CUDA, C++, and Python for efficient code development. Sparse matrix compression was used to minimize the memory usage and operation was fully vectorized to speed up computation. Memory usage was further reduced by using dynamic voxel spacing for various organs-at-risk. A Windows GUI framework was used to make the use of the TPS user-friendly. MongoDB was used to handle the patient data. The TPS has been seamlessly implemented as an Eclipse™ (Varian Medical System, Palo Alto, CA) plugin via transmission control protocol (TCP) socket communication, which provides users with on-the-fly interaction during the optimization. Without the support of the aperture blocks, most patient plan doses can be finished within 2-3s[65]. And one treatment plan can be finished within a couple of minutes (0.1-0.3s per iteration) running on AMD EPYC™ 7543 (AMD, Santa Clara, CA) equipped with 4 NVIDIA Ampere A100s (NVIDIA, Santa Clara, CA). The TPS has been used as a 2$^{nd}$ dose check and linear energy transfer (LET) evaluation tool for all the patients treated since the opening of our proton center (about 4,500 patients)[42,68,77,78] and as an interplay effects[22,24,48,74,79] evaluation tool in the last 6 years for all patients with mobile tumors (about 450 patients).

## D. Validation of Enhanced VPMC

The enhanced VPMC was validated by comparing the calculated results with MCsquare and RayStation MC in the following three experimental setups. First, single proton beamlets of various energies were delivered to a homogenous water phantom. Second, a single energy layer composed of many proton beamlets was delivered to the homogenous water phantom. Third, treatment plan doses were calculated in patient geometries with inhomogeneities considered. Please note that all the aperture blocks used in this study were non-divergent aperture blocks.

### D.1 Single Beamlet Validation in Water Phantom

There were two machines involved in the PBSPT-based SRS with an aperture block, as shown in Figure 1(a), the one with a range shifter (APRS) and the other without a range shifter (AP). For both AP and APRS machines, 97 proton energies used in our proton machines were used ranging from 71.3 to 228.8 MeV. The aperture block opening was 1 cm for both machines. In the validation, all beamlets used were delivered along the beam axis direction with the same lateral location, i.e., at the boundary of the aperture block opening to consider the influence of the aperture block upon the dose of the beamlets.

For every energy used in both machines, two simulations were conducted by two different dose engines for comparison, VPMC and MCsquare, respectively. The number of particles to be simulated (VP in VPMC and primary proton in MCsquare) were appropriately selected to achieve <0.5% and comparable statistical uncertainties for both dose engines[65]. The dose grid resolution used in simulations was 1 mm in all three cardinal directions. The dose distributions calculated by VPMC and MCsquare were compared in pairs through 3D gamma analysis[80] with a criterion of 2%/2 mm and a threshold of 10%.

**D.2 Energy Layer Validation in Water Phantom**

Figure 1(b)(c) show the validations in water phantoms, using two different configurations (henceforth referred as ConfigA and ConfigB). In ConfigA (Figure 1(b)), the isocenter was 108 mm below the water surface. Proton beamlets of 147.0 MeV were applied along the beam axis direction through an aperture block of 30 mm thickness that was mounted by the Hitachi holder aligned to the beam central axis with its distal surface 15 mm upstream the isocenter. Four aperture blocks were used with ascending central opening of 1, 2, 3, and 4cm. In ConfigB (Figure 1(c)), a few modifications were made compared to ConfigA: (1) a range shifter of 45 mm WET was introduced along the beam central axis and upstream the aperture block with its distal surface 190 mm upstream the isocenter (thus there is a 10mm gap between the aperture block and the range shifter), (2) the beamlet energy was elevated from 147.0 to 161.5 MeV. For both configurations, the mono-energetic proton beamlets were set to have the same intensity and were uniformly distributed in a square pattern with a spot spacing of 2.5 mm to cover the entire opening and the associated 5 mm outer margin (i.e., forming a circular irradiation field).

For each simulation scenario (4 in both ConfigA and ConfigB), similar calculations and comparisons were done as in the previous subsection "Single Beamlet Validation in Water Phantom".

**D.3 Treatment Plan Validation in Patient Geometries**

Ten brain cancer patients with small and shallow clinical target volumes (CTVs) previously treated using the photon SRS or stereotactic body radiotherapy (SBRT) technique at our institution were selected to validate the enhanced VPMC in patient geometries with inhomogeneities considered. Proton SRS treatment plans were first generated for these 10 patients in RayStation using its fast MC dose engine and then forward dose calculations of the generated plans were done using VPMC and MCsquare. The proton treatment plan information of these 10 patients is listed in Table 1. All of the MC calculations performed by VPMC, MCsquare and RayStation MC achieved a low statistical uncertainty of <0.5% with a dose calculation resolution of 2 mm in all three cardinal directions, according to the recommendation from AAPM Task Group 101[81]. The dose distributions calculated by VPMC and MCsquare were compared to the dose distributions calculated by RayStation MC using 3D gamma analysis with criteria of 3%/3 mm, 3%/2 mm, and 2%/2 mm, respectively, and a threshold of 10% (Table 2). The 3D gamma pasting rates from VPMC and MCsquare were statistical compared using the Wilcoxon signed-rank test to get the statistical significance. A *p*-value<0.05 is considered to be statistically significant. The calculation time of each plan by VPMC and MCsquare were also recorded and compared.

Table 1. Treatment planning information of the 10 selected patients.

| Patient # | Prescription Dose (Gy [RBE]) | Number of Fractions | CTV Volume (cc) | Treatment Field Angle | Range Shifter | Block Effective Diameter (cm) |
|---|---|---|---|---|---|---|
| 1 | 38 | 5 | 3.3 | T180G80[a] | No | 2.4 |
|  |  |  |  | T0G123 | No | 2.4 |
|  |  |  |  | T180G90 | Yes | 2.4 |
| 2 | 27 | 3 | 4.1 | T180G65 | Yes | 2.5 |
|  |  |  |  | T180G90 | Yes | 2.6 |
| 3 | 20 | 1 | 4.5 | T180G120 | Yes | 2.9 |
| 4 | 20 | 1 | 6.7 | T0G70 | Yes | 2.7 |
| 5 | 20 | 1 | 0.4 | T180G90 | No | 1.8 |
|  |  |  |  | T0G90 | No | 1.8 |
| 6 | 25 | 5 | 10.2 | T180G45 | Yes | 3.6 |
|  |  |  |  | T180G90 | Yes | 3.4 |
|  |  |  |  | T180G180 | No | 3.6 |
| 7 | 39 | 5 | 7.4 | T180G180 | Yes | 3.0 |
| 8 | 12 | 1 | 43.3 | T250G90 | No | 5.2 |
|  |  |  |  | T310G70 | No | 5.1 |
|  |  |  |  | T0G105 | No | 5.2 |
|  |  |  |  | T270G355 | No | 4.6 |
| 9 | 24 | 3 | 1.3 | T180G80 | Yes | 1.8 |
| 10 | 16 | 1 | 2.6 | T180G160 | Yes | 2.6 |
|  |  |  |  | T0G160 | Yes | 2.6 |
| Mean | - | - | 8.38 | - | - | 3.11 |
| SD | - | - | 12.62 | - | - | 1.11 |

*abbreviations*: CTV for clinical target volume, SD for standard deviation, RBE for relative biological effectiveness.

[a] T for table and G for gantry.

### E. Robust Treatment Planning

Three typical patients, patient 3 (SRS with a range shifter), patient 5 (SRS without a range shifter), and patient 7 (5 fraction SBRT with a range shifter) were selected for robust optimization using the enhanced VPMC. Our institutional guidelines on the dose volume constraints (DVCs) for PBSPT-based SRS/SBRT in the treatment of brain cancers, based on the version for photon-based SRS/SBRT in the treatment of brain cancer, was used and listed in Table 3. Only the DVCs for targets were listed since the DVCs for OARs were easily met for the three selected patients. CI$x$%

⩽*y1~y2* is the conformity index (CI) constraint, where CI*x*% was defined as the ratio of the volume enclosed by the *x*% iso-dose (*x*% of the prescription dose) to the volume of the structure of interest (SOI). This CI constraint has a soft upper limit (better below) *y1* and a hard limit (must below) *y2*. Similarly, RHI*x*%⩽*y1~y2* is the radical dose homogeneity index (RHI) with RHI*x*% defined as the ratio of the maximum dose within the SOI to the *x*% of the prescription dose.

For robust optimization, the proton range uncertainty was modelled by scaling the CT's relative-to-water-stopping-power-ratio by ±3%. The patient setup uncertainty was modelled by shifting the isocenter positively and negatively in all three cardinal directions with a magnitude of 2 mm per out institutional protocol. In total, 13 uncertainty scenarios were included in the robust optimization. Per our institutional protocol, single-field optimization (SFO) was the preferred method for robust optimization. Alternative multiple-field optimization (MFO) would be used if SFO failed to meet the clinical requirements.

To generate the patient-specific aperture opening edge, a 1 mm margin was first added to the CTV to form the planning target volume (PTV). Then PTV which was further expanded with a 2 mm margin to form the patient-specific aperture opening edge to have good target coverage.

**Results**

**A. Single Beamlet Validation in Water Phantom**

Three typical proton energies (low, medium, and high) were selected to illustrate the results: 82.0, 140.2, and 228.8MeV (Figure S-1). For both machines (AP and APRS), the percentage depth dose (PDD) curve and the integrated depth dose (IDD) curve from dose distributions calculated by VPMC and MCsquare were compared and found to be in excellent agreement. The 3D gamma passing rates were 99.55% on average with the minimum of 98.78%, under the 2%/2mm/10% criterion (Table S-1).

**B. Energy Layer Validation in Water Phantom**

The IDD curves (top row in Figure 2), log PDD curves (middle row in Figure 2), and iso-dose contours in the transverse plane (20% and 80% of the maximum dose, bottom row in Figure 2) were compared between the doses distributions calculated by MCsquare and VPMC for ConfigA (see Figure S-2 for ConfigB). In Figure 2, from left to right, each column corrsponded to an aperture block opening size, 1, 2, 3, and 4 cm, respectively. Results calculated by VPMC agreed well with the results calculated by MCsquare (curves and contours are largely overlapped). In the 3D gamma analysis, the average passing rate of all validation scenarios (ConfigA and ConfigB) was 99.71±0.23% (Table S-2).

**C. Treatment Plan Dose Validation in Patient Geometries**

Table 2 showed the 3D Gamma passing rates by comparing the dose distributions calculated by MCsquare and VPMC to those calculated by RayStation MC, respectively, which were 98.31±1.62% and 98.47±1.63% with the criterion of 3%/3mm/10% (the former was MCsquare vs. RayStation MC while the latter was VPMC vs. RayStation MC hereafter), 97.78±1.97% and 97.79±2.21% with the criterion of 3%/2mm/10%, and 94.57±3.02% and 94.76±3.96% with the criterion of 3%/2mm/10%. No statistically significant differences of the 3D Gamma passing rates were found between the results calculated by MCsquare and by VPMC ($p$=0.77, 0.70, and 1.0, respectively), indicating a comparable simulation accuracy between VPMC and MCsquare. The calculation time was drastically decreased from 112.45±114.08 seconds when using MCsquare to 8.20±6.42 seconds when using VPMC.

Two representative patients (patient 5 without range shifter and patient 7 with range shifter) were selected for comparison of the dose distributions calculated by VPMC, MCsquare, and RayStation MC (Figure 3[a]-[c] for patient 5 and Figure 3[d]-[f] for patient 7). The dose profiles longitudinal (Figure 3[b] [e]) and perpendicular (Figure 3[c] [f]) to the beam axis in the target region agreed well among VPMC, MCsquare, and RayStation MC.

Table 2. 3D gamma passing rates (%) of treatment plan dose validation in patient geometries by comparing MCsquare and VPMC with RayStation MC with criteria of 3%/3mm/10%, 3%/2mm/10%, and 2%/2mm/10%, respectively.

| Patient # | $\gamma$-pass rate (%) 3%/3mm | | $\gamma$-pass rate (%) 3%/2mm | | $\gamma$-pass rate (%) 2%/2mm | | time (s) | |
|---|---|---|---|---|---|---|---|---|
| | MC2 | VPMC | MC2 | VPMC | MC2 | VPMC | MC2 | VPMC |
| 1 | 98.01 | 99.87 | 97.34 | 99.83 | 95.37 | 98.89 | 163.54 | 10.59 |
| 2 | 99.65 | 99.21 | 99.44 | 98.76 | 97.30 | 95.90 | 46.48 | 4.36 |
| 3 | 99.46 | 96.41 | 98.72 | 94.10 | 93.91 | 87.13 | 78.32 | 8.42 |
| 4 | 98.61 | 98.73 | 98.16 | 98.36 | 94.06 | 94.72 | 82.07 | 7.57 |
| 5 | 99.31 | 99.36 | 99.13 | 99.02 | 97.40 | 97.10 | 84.98 | 3.59 |
| 6 | 95.06 | 99.61 | 93.34 | 99.23 | 89.61 | 97.23 | 93.86 | 4.44 |
| 7 | 96.84 | 96.17 | 95.96 | 95.22 | 91.14 | 91.04 | 80.07 | 6.74 |
| 8 | 99.85 | 99.79 | 99.75 | 99.65 | 98.75 | 98.59 | 419.16 | 25.16 |
| 9 | 96.72 | 95.90 | 96.95 | 94.69 | 91.77 | 90.24 | 18.23 | 3.25 |
| 10 | 99.60 | 99.60 | 99.05 | 99.07 | 96.41 | 96.74 | 57.82 | 7.91 |
| Mean | 98.31 | 98.47 | 97.78 | 97.79 | 94.57 | 94.76 | 112.45 | 8.20 |
| SD | 1.62 | 1.63 | 1.97 | 2.21 | 3.02 | 3.96 | 114.08 | 6.42 |
| *P*-value | 0.77 | | 0.70 | | 1.0 | | - | |

*abbreviations*: MC2 for MCsquare, SD for standard deviation.

**D. Robust Optimization**

The robustly optimized plans using the enhanced VPMC dose engine for the 3 representative patients all met the DVCs of the targets per our institutional protocol (Table 3). In contrast, the robustly optimized plans using the original VPMC (without modelling the aperture block) when adopting the same spot arrangements and DVCs in the robust optimization, didn't always meet the DVCs of the targets (Table 3). The mean calculation time for 13 influence matrices in robust optimization by VPMC was 41.6 seconds and the subsequent on-the-fly "trial-and-error" optimization procedure took 71.4 seconds on average for the selected three patients.

Table 3. Target DVH indices of the robustly optimized plans for the three representative patients.

|  | Patient 3 | Patient 5 | Patient 7 |
|---|---|---|---|
| V100% (%, ≥ 99%) | 99.03 | 100.00 | 99.82 |
| $D_{min}$ (%, ≥ 90%) | 97.65 | 100.13 | 94.49 |
| CI100% (≤ 1.5~2.0) | 1.94 | 1.73 | 1.96 |
| RHI100% (≤ 1.3~1.4) | 1.15 | 1.13 | 1.19 |
| IM calculation time (s) | 48.9 | 27.4 | 48.5 |
| Optimization time (s) | 80.1 | 40.8 | 93.3 |

*abbreviations*: CI for conformity index, RHI for radical dose homogeneity index, DVC for dose-volume constraint

**Discussion**

In this work, we enhanced a fast proton dose engine, VPMC[65], to support the aperture block to achieve fast MC-based dose calculation (within seconds) and robust optimization (within a couple of minutes) for PBSPT-based SRS to treat brain cancer. The aperture block has been proposed to sharpen the target dose penumbra in PBSPT-based SRS to better protect the nearby normal tissues. The crossing number algorithm[42] was adopted to model the aperture block, which led to more accurate modelling of the aperture block opening for proton dose calculation.

In the validations in water phantoms, some mismatches between the IDD curves calculated by VPMC and MCsquare were observed in the entrance region, especially for smaller aperture openings (Figure 2[a] and Figure S-2[a]). We didn't pay much attention to such mismatches because we assumed mismatches of such level of magnitude (the larges mismatch with 1 cm diameter was less than 5% compared to the Bragg peak dose) in the entrance region would have limited impact clinically. Possible reasons for such mismatches could be the different methods to model the opening of apertures and the slit scattered protons among RayStation MC, VPMC and MCsquare. Further research to investigate this small difference will be done in the future.

Overall, the 3D Gamma passing rates for the validations in water phantoms outperformed those in the patient geometries. There are two possible reasons: first, the heterogeneities in patient geometries might introduce uncertainties in the dose calculation; second, the dose calculation resolution used for validation in the water phantoms was half of the one used for validation in the patient geometries (1 mm vs. 2 mm). Given that the target volumes were rather small (mean: 8.32 cc and 4.50 cc when excluding the largest one with a volume of 43.3 cc) in the selected patients, the routinely used dose calculation resolution (2~3 mm) might not be small enough to calculate doses for the selected patients accurately. In order to understand the possible impact of the dose calculation resolution upon the final dose distributions calculated by MCsquare and VPMC, we had conducted the same dose calculations for the selected 10 patients with a dose calculation resolution of 2mm and 1mm, respectively. An obvious increase of the 3D Gamma passing rates was observed when changing the dose resolution from 2mm to 1mm, especially for the 2%/2mm criterion where the 3D Gamma passing rate was increased from 96.94±2.82% to 99.05±1.10% (*P*-value = 0.002) (Table S-3). This suggested that in the PBSPT-based SRS treatment for small tumors, the dose calculation resolution should be cautiously selected (the smaller the better) for the treatment plan optimization and dose calculation.

This study has certain shortcomings. First, the aperture block used in this study is that it is non-divergent, which could lead to collimator contamination problem by the edge-scattered protons, especially for shallow tumors with adjacent lateral and proximal OARs. A divergent aperture block could help mitigate this problem[40,82], which will be our next step to further enhance the benefits of the use of the aperture block in the PBSPT-based SRS treatment for brain cancer patients with shallow small targets. Second, patient-specific aperture used in this study was fabricated according to the largest tumor cross-section (plus some margin) from the beam eye view, thus only improving the dose lateral penumbra at the depth with the largest tumor cross-section. In the future, we will investigate the feasibility of an energy-layer-specific dynamic aperture to cover the whole target in all depths, i.e., the whole target in the beam direction.

**Conclusion**

In this work, we have enhanced VPMC to model aperture blocks used in the PBSPT-based SRS in the treatment of brain cancer patients with small targets. The enhancement includes both fast MC-based dose calculation (average 8.2 seconds) and fast robust optimization (average 113.0 seconds). The benchmark results against MCsquare in water phantoms and against MCsquare/RayStation MC in patient geometries are excellent. Our enhanced VPMC can be used to robustly optimize a PBSPT-based SRS plan efficiently and accurately.

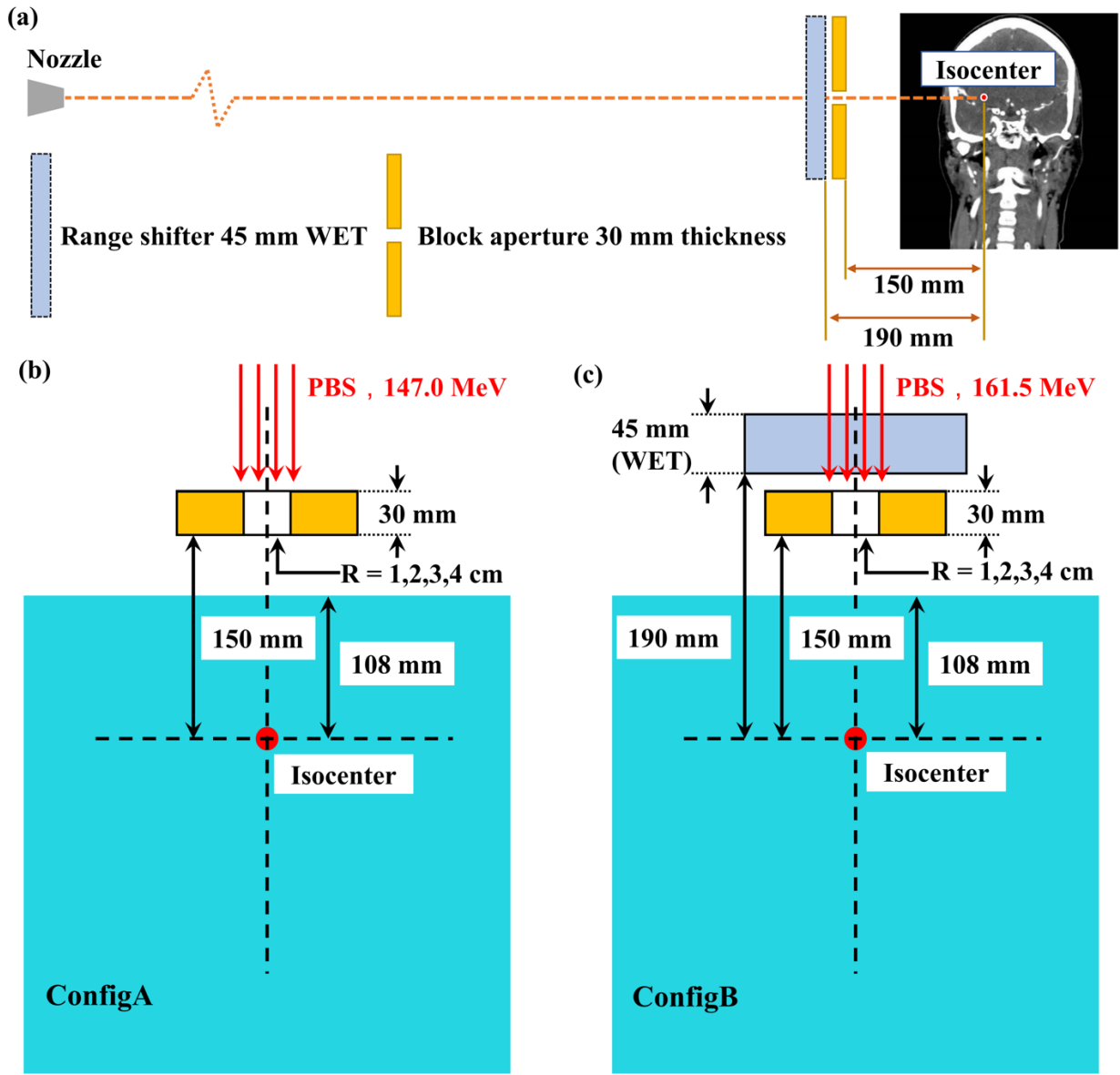

Figure 1. Aperture block and range shifter configurations at our institution and aperture block w/ and w/o range shifter configurations used in the water phantom validation. (a) Aperture block and range shifter configurations in PBSPT-based SRS. (b) Configurations of the aperture block without range shifters in ConfigA. (c) Configurations of the aperture block with a range shifter in ConfigB.

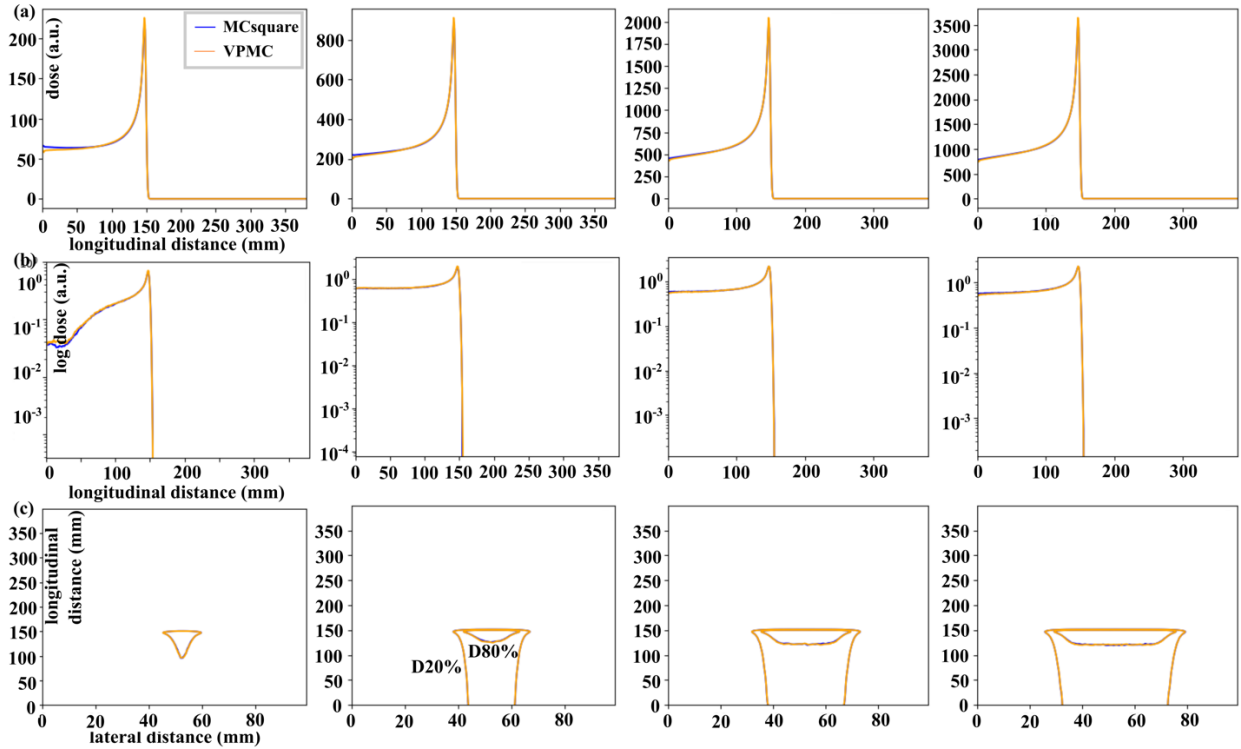

Figure 2. Results of the energy layer validation in homogenous water phantoms with ConfigA. Row (a) are the IDD curves, (b) log PDD curves, and (c) contours of iso-dose contours (20% and 80% of the maximum dose) in the transverse planes. From left to right, columns display the results calculated with aperture openings from 1cm to 4cm, respectively. Blue lines are the results calculated by MCsquare, while orange lines are the results calculated by VPMC.

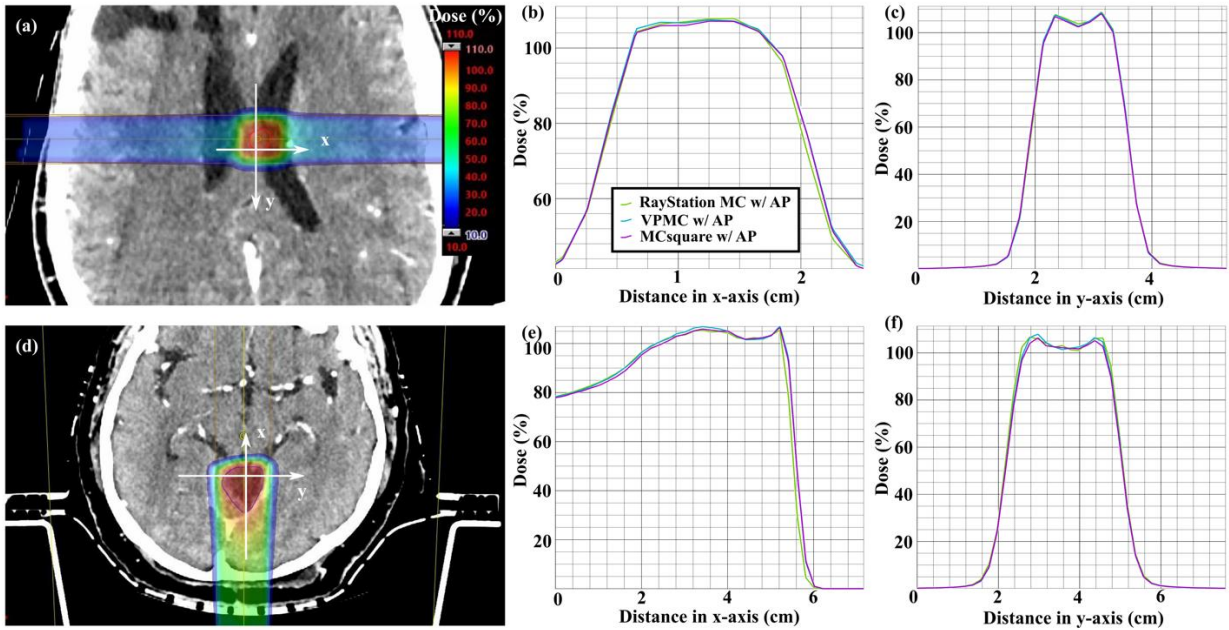

Figure 3. Dose distributions on a transverse plane of patient 5 ((a)-(c)) and patient 7 ((d)-(f)) calculated by VPMC. Red contours are CTVs. Dose profiles are compared in (b) (e) longitudinal direction (horizontal x arrows in (a) and (d)) and (c) (f) lateral direction (vertical y arrows in (a) and (d)). Results with the aperture block from RayStation MC, VPMC, and MCsquare are in green, blue and purple, respectively.